\documentclass[useAMS,usenatbib]{mn2e}
\usepackage{graphics}

\def\oversim#1#2{\lower0.5pt\vbox{\baselineskip0pt \lineskip-0.5pt
     \ialign{$\mathsurround0pt #1\hfil##\hfil$\crcr#2\crcr\sim\crcr}}}

\begin{document}
  \title[ A cometary knot in the Helix Nebula]
  {VLT / Infrared Integral Field Spectrometer Observations of
Molecular Hydrogen Lines in the Knots in the Planetary Nebula NGC 7293 (the Helix Nebula)
 \thanks{Based on observations with European Southern Observatory,
 Very Large Telescope with an instrument, SINFONI
(the proposal numbers: 076.D-0807).}
}
\author[M. Matsuura et al.]
{M.~Matsuura$^{1,2,3}$, 
A.K.~Speck$^{4}$,
M.D.~Smith$^{5}$,
A.A.~Zijlstra$^{3,6}$,
S.~Viti$^{7}$,
\newauthor
K.T.E.~Lowe$^{8}$, 
M.~Redman$^{9}$,
C.J.~Wareing$^{3}$,
E.~Lagadec$^{3}$,
\\
$^{1}$ National Astronomical Observatory of Japan, Osawa 2-21-1, 
	Mitaka, Tokyo 181-8588, Japan \\
$^{2}$ School of Mathematics and Physics, 
	Queen's University of Belfast, Belfast BT7 1NN, Northern Ireland, United Kingdom \\
$^{3}$  School of Physics and Astronomy, University of Manchester,
	Sackville Street, P.O. Box 88, Manchester M60 1QD, United Kingdom \\
$^{4}$ Physics and Astronomy, University of Missouri, Columbia, MO 65211, USA \\
$^{5}$  Centre for Astrophysics and Planetary Science,  School of Physical Sciences,
	The University of Kent, Canterbury CT2 7NH, United Kingdom \\
$^{6}$	South African Astronomical Observatory, P.O.Box 9, 7935     
        Observatory, South Africa \\
$^{7}$ Department of Physics and Astronomy, University College London, 
	Gower Street, London WC1E 6BT, United Kingdom \\
$^{8}$  Science and Technology Research Centre, University of Hertfordshire, 
	College Lane, Hatfield, Hertfordshire AL10 9AB, United Kingdom \\
$^{9}$ Department of Physics, National University of Ireland Galway, Galway, Republic of Ireland \\
}

\date{Accepted. Received; in original form }
\pagerange{\pageref{firstpage}--\pageref{lastpage}} \pubyear{2007}

\maketitle
\label{firstpage}

\begin{abstract}

  Knots are commonly found in nearby planetary nebulae (PNe) and star forming
 regions. Within PNe, knots are often found to be associated with the
 brightest parts of the nebulae and understanding the physics involved in
 knots may reveal the processes dominating in PNe.  As one of the closest PNe,
 the Helix Nebula (NGC 7293) is an ideal target to study such small-scale
 ($\sim$300\,AU) structures.  We have obtained infrared integral spectroscopy
 of a comet-shaped knot in the Helix Nebula using SINFONI on the Very Large
 Telescope at high spatial resolution (50--125~mas).  With spatially
 resolved 2\,$\mu$m spectra, we find that the H$_2$ rotational temperature
 within the cometary knots is uniform.  The rotational-vibrational
 temperature of the cometary
 knot (situated in the innermost region of the nebula, 2.5~arcmin away from
 the central star), is  1800~K, higher than the temperature seen in
 the outer regions (5--6\,arcmin from the central star) of the nebula (900~K), showing that the
 excitation temperature varies across the nebula.  The obtained intensities
 are reasonably well fitted with 27~km\,s$^{-1}$ C-type shock model.
 This ambient gas velocity is slightly higher than the observed [He{\small II}] wind velocity of 13~km\,s$^{-1}$.
 The gas excitation can also be reproduced with a PDR (photo dominant region) model, but this requires
 an order of magnitude higher UV radiation. Both models have limitations,
 highlighting the need for models  that treats both hydrodynamical physics and the PDR.
\end{abstract}


%
\large

\maketitle


\section{Introduction}

In recent years it has become clear that knots of dense material are common in
nebulae, including Planetary Nebulae \citep[PNe; e.g.][]{Odell02} and
star-forming regions \citep[e.g.][]{McCaughrean97}.  In the one of the best studied PN
case, that of the Helix Nebula (NGC 7293), knots in the inner regions have a
comet-like shape \citep{Odell96} and are thus known as cometary knots.  The
Helix nebula is estimated to contain more than 20,000 cometary-shaped knots
\citep{Meixner05}.  The apparent commonality of occurrence in PNe of these
knots has lead to the assertion that all circumstellar nebulae are clumpy in
structure \citep[e.g.][]{Odell02, Speck02, Matsuura05b}.  As knots occupy the
brightest parts of the planetary nebulae, understanding their physical nature
is essential to understanding  the dominant physics governing the  nebula.

The origin of these knots remains unknown. When the knots form is also
disputed: they may have been present during the preceding AGB (Asymptotic
Giant Branch) phase \citep[e.g.][]{Dyson89}, or only have formed during the PN
phase.  The suggested formation mechanisms fall into two main scenarios.
Firstly, photo-evaporation from condensations which pre-exist in the
circumstellar envelope were considered.  \citet{Speck02} suggest that the UV
radiation from the central star heats the surface of a condensation, emitting
H$\alpha$ and H$_2$ on the facing side. 
 Secondly, the interaction of a fast stellar wind with the
slowly expanding AGB wind, during the early stage of PN phase, has been
considered \citep[e.g][]{Vishniac94, Pittard05, GS06}.
\citet{Pittard05} found that cometary knots may be formed through instabilities
where a supersonic wind impacts a subsonic wind.

In addition to being seen as inhomogeneities within the ionized gas emission
from nebulae, the knots also contain molecular gas \citep{Speck02,
Speck03}. This yields a potential clue to their origin. \citet{Redman03}
argued that some molecular species formed in the AGB atmosphere can survive if
molecules are shielded in clumps from UV radiations by extinction.  But if the
knots (re-)form during the ionized PN phase, the earlier AGB molecules were
photo-dissociated, and the molecules reform within the knots, under conditions
of relatively low density and higher UV intensity compared to the AGB wind. In
these conditions only simple molecules are expected to be present
\citep{Wood04}.

Using seeing-limited observations (1.2~arcsec) of the H$_2$ v=1--0 S(1) line
in a single knot, \citet{Huggins02} found H$_2$ gas in both the head and the
tail of the knot, following the distribution of ionised gas traced by
H$\alpha$ and [N{\small II}].  
In contrast, CO thermal emission was found to
emanate only from the tail.  
Higher angular resolution images obtained with
HST \citep{Meixner05} resolved the globular H$_2$ emission into
crescent-shaped regions.  With these previous studies, it is not clear whether
the hydrogen molecules and ionised gas are co-located within the head or
spatially separated.

The excitation mechanisms for molecular hydrogen in planetary nebula have long
been controversial, with contention between fluorescent/thermal excitation in
photon-dominant regions (PDR) or shock excitation.  \citet{Tielens85} and
\citet{Black87} showed that vibrationally excited H$_2$ (by FUV pumping)
traces the surface of dense regions in the process of becoming photo-ionised.
On the other hand, molecules can form in
a post-shock region \citep{Neufeld89}, and H$_2$ can be excited by shocks
\citep{Beckwith80, Hollenbach89}.  The cooling region after the shock can be
resolved if it is C-type (continuous) shock, but the region is too small to be
resolved for J-type (jump) shocks.  A comparison of models with spatially
resolved spectra, which covers multiple line ratios, will help understanding
the excitation mechanism of H$_2$.

The Helix Nebula (NGC\,7293) is one of the nearest PNe, with a largest
diameter of more than half a degree \citep{Hora06} and a parallax distance of
219\,pc \citep{Harris07}.  Because of its distance, small-scale structures
inside the nebula are well resolved and this nebula is used as a proxy to
understand the structures found in PNe.

We have observed a cometary knot in the Helix Nebula, using the
adaptive-optics-assisted, near-infrared integral field spectrometer on the
Very Large Telescope. We have achieved 50--100~mas spatial resolution. Our
data provide the first spatially resolved spectra within a knot at 2~$\mu$m and
further this is the highest spatial resolution image+spectra of this PN at
this wavelength.  Our observations of H$_2$ spectral line ratios
allow us to derive the excitation temperatures within the knot, as well as
differentiation between the possible excitation mechanisms
of H$_2$.

\section{Observations and Analysis}


\begin{table*}
 \centering
 \begin{minipage}{140mm}
  \caption{Observing log \label{table-targets}}
 \begin{tabular}{lcccllccccc}
  \hline
Target &  & & & &  \multicolumn{3}{l}{Telluric Standard} \\
 &  & Pixel scale of the camera & Pixel scale in reduced data & Exp$^{\dagger}$  & Name & Spec & K-mag$^{\ddag}$ \\
 & &  mas$^2$&  mas$^2$ & \\
 \hline
K1   &  & 125$\times$250& 125$\times$125 & 300s$\times$10 & Hip~115329 &G2V & 7.552$\pm$0.017 \\
& &  & & 600s$\times$10   \\
& & 50$\times$100& 50$\times$50 & 600s$\times$10&  Hip~023422 & G2V &  7.862$\pm$0.017\\
\hline
\end{tabular}
\footnotetext{$^{\dagger}$: Exp: Exposure time. }
\footnotetext{$^{\ddag}$: K-mag: K'-band magnitudes from 2MASS }
\end{minipage}
\end{table*}

\begin{figure}
\centering
\resizebox{\hsize}{!}{\includegraphics{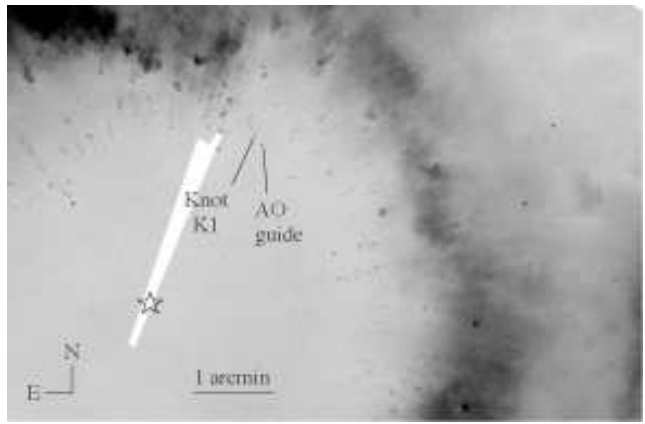}}
\caption{ The location of the cometary knot K1 is plotted on the F658N
([N{\small{II}}]+H$\alpha$) image \citep{Odell04}, together with the AO guide
star. The star symbol indicates the place of the central star.
\label{Fig:target}
}
\end{figure}

A knot in the Helix Nebula was observed by the Spectrograph for INtegral Field
Observations in the Near Infrared \citep[SINFONI;][]{Eisenhauer03, Bonnet04}
installed at the Cassegrain focus of the Very Large Telescope.  The grating
for the K-band was used.  The wavelength band of the grating was mapped to the
2048 pixel detector in the dispersion direction. We used two plate-scales: the
spatial resolutions are 125$\times$250~mas$^2$ and 50$\times$100~mas$^2$.  For
all of the observations, a nearby star (RA=22:29:33.017 Dec=$-$20:48:12.68)
was used as Adaptive Optics (AO) guide star.  The spatial resolution is
usually determined by the pixel scale while using the 125$\times$250~mas$^2$
plate-scale (i.e. the images are undersampled), while for the
50$\times$100~mas$^2$ scale both the corrections of the AO system and the
pixel scale are important.  The spectral resolutions $\lambda / \Delta
\lambda$ are 4490 and 5090 for the 125$\times$250~mas$^2$ and
50$\times$100~mas$^2$ cameras, respectively.  The observing log is summarized
in Table~\ref{table-targets}.  The coordinates of our target knot K1 are
22:29:33.414, $-$20:48:04.73 (J2000) measured from the data used by
\citet{Odell04}.  This knot is noted as ID 6 in \citet{Meaburn98} and is about
2.5\,arcmin away from the central star (Fig.\,\ref{Fig:target}). It is located at
the inner rim of the ring-shaped nebula filled with cometary knots.

The observing run was carried out on the first-half nights from 2$^{\rm nd}$
to 4$^{\rm th}$ of November 2005.  The sky was clear on Nov 3, with occasional
thin cloud passing on November 2 and 4.  The telluric standards were observed
immediately after the target.  The telluric standard stars were used for flux
and spectral-response calibration, as well as for the measurements of the
point spread function.  Model spectra from \citet{Pickles98} are used as
template spectra for the telluric standard stars.  The wavelength resolution of
the model spectra is 4000 only, but we assume that this difference in wavelength
resolution is not critical for our analysis of the emission lines.

The field of view was slightly jittered to minimize the influence of bad
pixels.  The sky level was measured at +10~arcmin away in the declination
direction, which is outside of the bright part of the nebula.  Some residual
of the sky level was left, and we use the integral field to remove the
residual sky level.

The ESO data reduction pipeline for SINFONI on GASGANO \citep{Modigliani07}
was used.  Distortion correction and flat correction were adopted.  Wavelength
calibration used the Ne and Ar wavelength lamps, linearly interpolated
throughout the entire wavelength coverage.  After the data reduction with
GASGANO, the final pixel was re-sampled to a 125 and 50 mas scale for
125$\times$250~mas$^2$ and 50$\times$100~mas$^2$ plate scale images,
respectively.

The absolute flux calibration has some uncertainty, due to the possibility of
occasional thin cirrus (up to 0.2\,mag) and the generic difficulty in the
calibration of diffuse intensity using an AO-assisted point source.  We
reduced two 125 \,mas target spectra using two different exposure time
independently, and the final spectra and images were obtained by average of
these two data sets.  The average of our measured flux over
2$\times$2\,arcsec$^{-2}$ is $1.2\times10^{-4}$
erg\,s$^{-1}$\,cm$^{-2}$\,sr$^{-1}$.  \citet{Speck02} gives the intensity of
K1 at H$_2$ v=1--0 S(1) as $\sim5$--$10\times10^{-5}$
erg\,s$^{-1}$\,cm$^{-2}$\,sr$^{-1}$; the knots are not fully resolved
(2\,arcsec pixel$^{-1}$) and the flux of this knot is close to the detection
limit ($\sim1\times10^{-4}$erg\,s$^{-1}$\,cm$^{-2}$\,sr$^{-1}$).  A knot in
the outer regions of the Helix, with a similar flux level to our knot in
\citet{Speck02}'s measurements, was observed by \citet{Meixner05}. They find a
surface brightness of H$_2$ of $\sim 1\times10^{-4}$
erg\,s$^{-1}$\,cm$^{-2}$\,sr$^{-1}$. We conservatively adopt a systematic
calibration uncertainty of 50 percent.  Further, the sky condition affects the
correction of the atmospheric transmittance, especially shortward of 2\,$\mu$m
and longward of 2.4\,$\mu$m, due to the variation of terrestrial H$_2$O.  The
relative intensities of lines below 2.0\,$\mu$m and above 2.4\,$\mu$m have an
error of $\sim$30\%.  We ignore the extinction at 2\,$\mu$m; as discussed
later the extinction affects the absolute intensity less than 5\,\%
($A_{\lambda=2.128\mu \rm{m}} =0.05$\,mag).

\section{Description of the data}

\subsection{Morphology}

\begin{figure*}
\centering
\resizebox{\hsize}{!}{\includegraphics{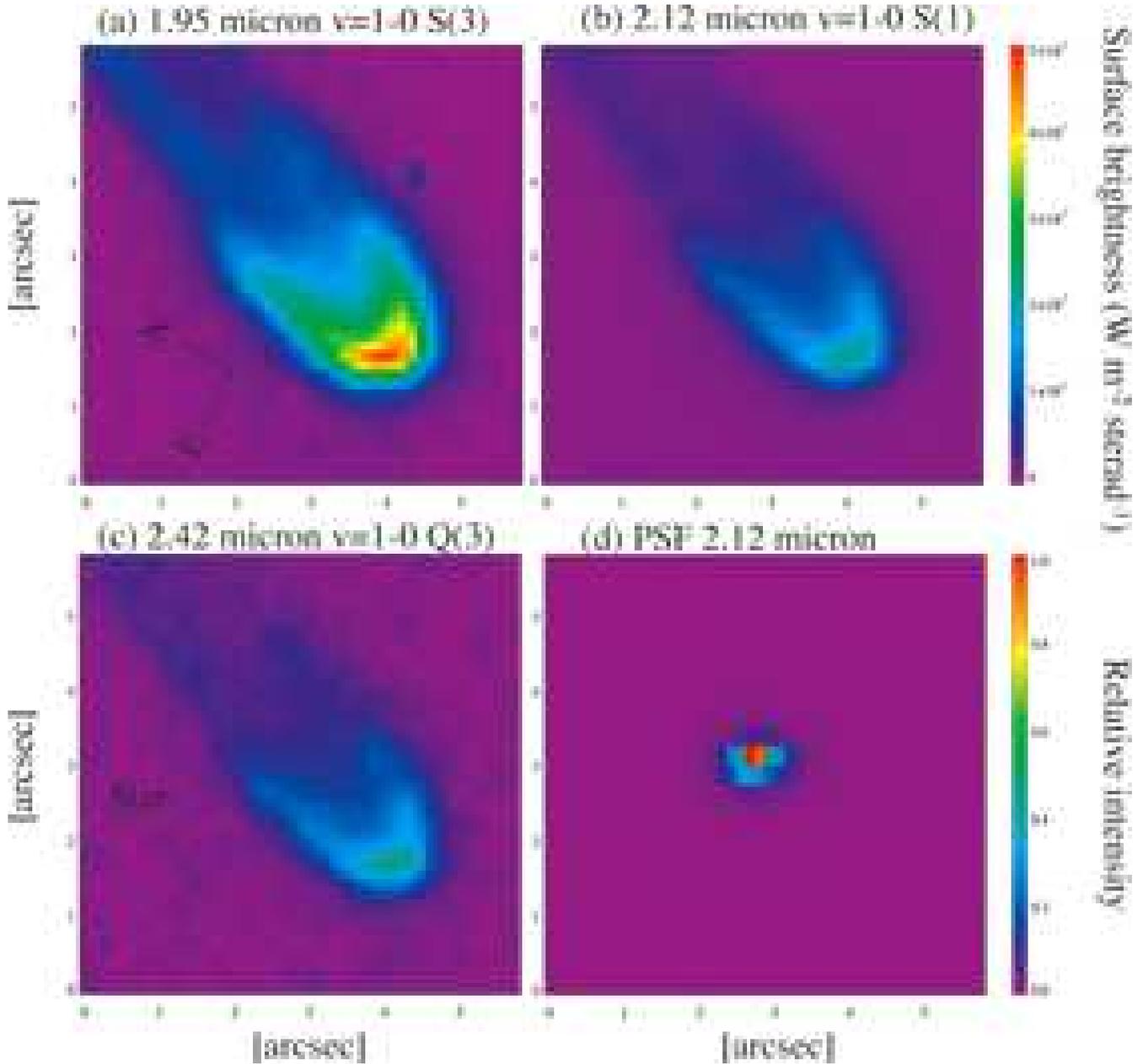}}
\caption{
Images of the cometary knot K1, taken by 250 mas camera.
The PSF at 2.12~$\mu$m is found in (d).
The direction of the central star is marked in (c)
\label{Fig:image_p5g1_250mas}
}
\end{figure*}
\begin{figure*}
\centering
\resizebox{\hsize}{!}{\includegraphics{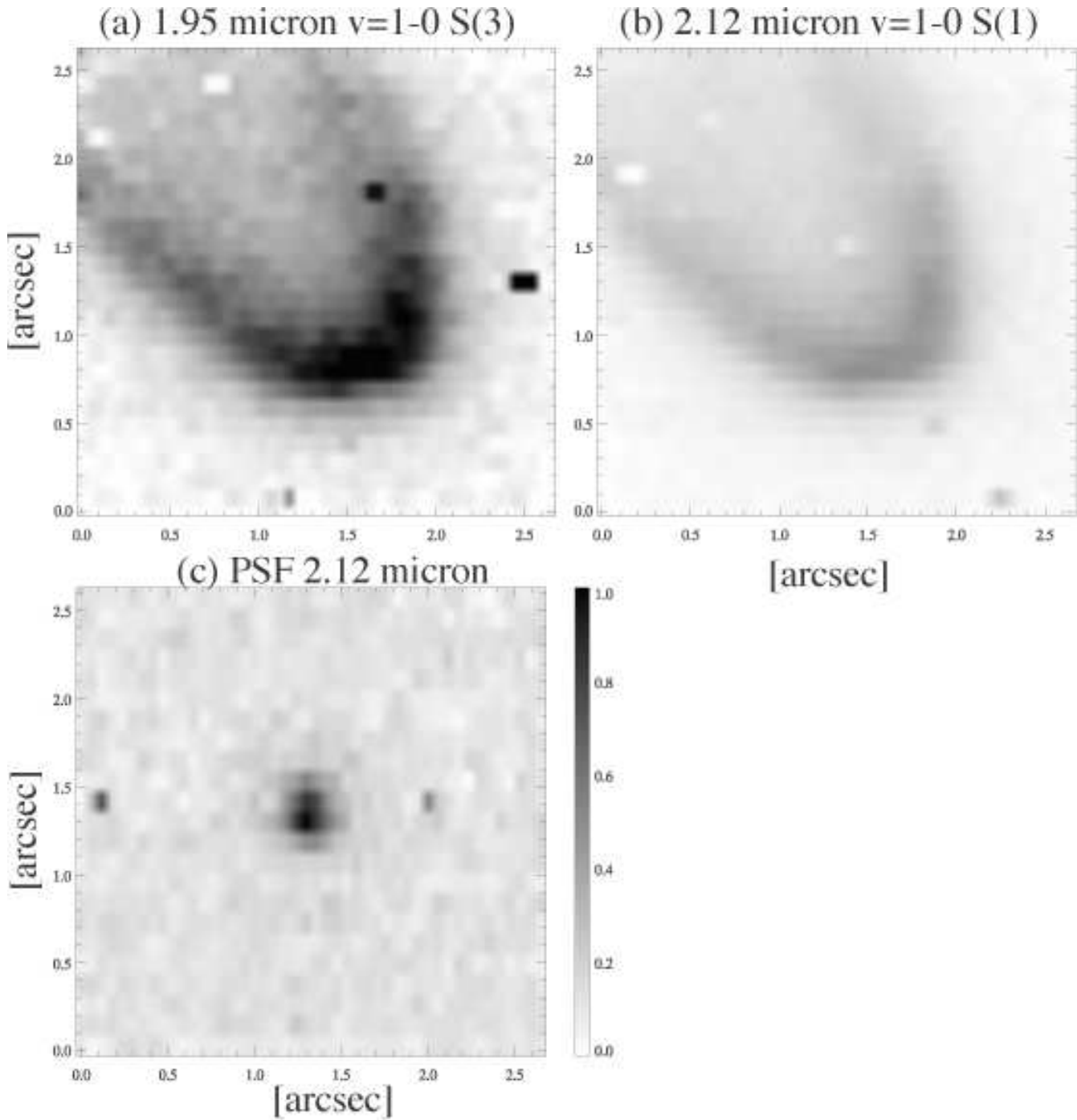}}
\caption{
Image of the cometary knot K1, taken with the  100 mas camera.
The PSF at 2.12~$\mu$m is found in (c).
\label{Fig:image_p5g1_100mas}
}
\end{figure*}

\begin{figure}
\centering
\resizebox{\hsize}{!}{{\includegraphics*{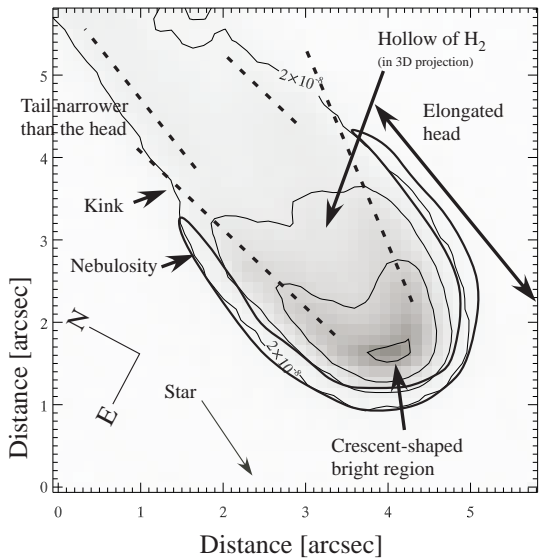}}}
\caption{ Annotated 2.12\,$\mu$m image of the knot K1 (see text for details).}
\label{Fig:shape}
\end{figure}
\begin{figure}
\centering
\resizebox{\hsize}{!}{\rotatebox{90}{\includegraphics*[66,157][536,717]{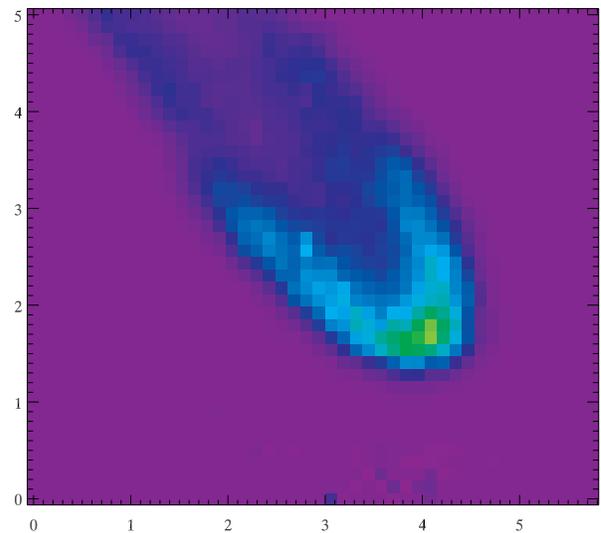}}}
\caption{ Deconvolved image at 2.12 $\mu$m. A fain emitting region at the 
rim near the crescent is found, as well as narrowing of the tail.}
\label{Fig:deconvolved}
\end{figure}
\begin{figure}
\centering
\resizebox{\hsize}{!}{\includegraphics*[46,0][430,560]{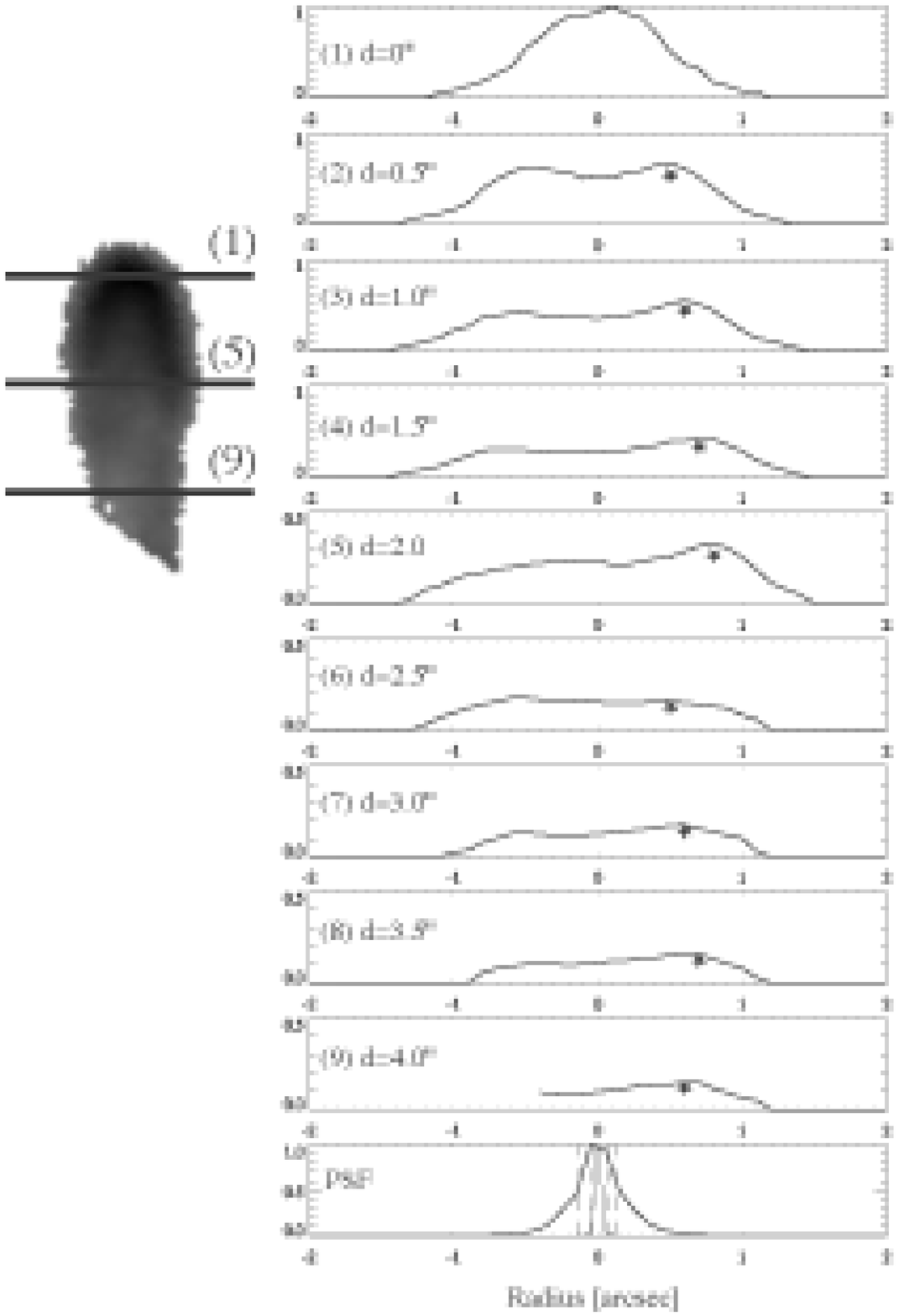}}
\caption{ Cross cut of Fig.~\ref{Fig:image_p5g1_250mas} along the radial axis,
whose intensity is scaled to the maximum intensity of the knot.
Approximate locations of cuts are indicated on the left panel.
The data are smoothed with 250 mas bin along long axis of the knot.
The intensity peaks on east side (right side in this figure) of the knot are marked with stars
in each panel.
The bottom panel shows the cross cut of the telluric standard (bold line), 
airy disk for 8.2 meter circular aperture (thin solid line),
and pixel size (125 and 250 mas; thin dash lines).
The spatial resolution is determined by the pixel scale (sampling rate).
The star marks traces that the head is widen until at distance ($d$)=2.0\,arcsec 
and then narrowing follows $d>$2.5.
Tail of the PSF is less than 10\,\%  of the peak at 0.5\,arcsec.
}
\label{Fig:sinfoni-cross}
\end{figure}

Fig.~\ref{Fig:image_p5g1_250mas} shows the image of the cometary knot K1 as
seen in the three strongest H$_2$ lines.  The knot shows an elongated head 
with a narrower tail.  The brightest emission is found in a crescent near the
tip of the head.  The crescent ends in two linear segments, indicated by the
dotted lines in Fig.~\ref{Fig:shape} on  the 2.12 $\mu$m v=1--0
S(1) image.  These segments are not co-aligned and deviate from the direction
of the narrower tail. Overall, this gives the impression of a `tadpole' shape,
as opposed to the cylindrical shape favoured by \citet{Odell96}, although
in either case the overall shape is largely axisymmetric. Both the 
linear segments and the tail are brighter on the eastern side of the knot.

The peak emission is located slightly behind the tip (Fig.~\ref{Fig:shape}).  There is 
a faint nebulosity around the bright head.  This faint nebulosity appears not
to be the wing of the point spread function (PSF). 
The faint halo extends along the linear segments where the peak emission is much
fainter.  
The higher resolution image (50~mas per pixel) in Fig.~\ref{Fig:image_p5g1_100mas} also
shows this faint nebulosity, suggesting
that at least part of this faint nebulosity is real.

The apparent diameter at the head is about 2.5 arcsec including the faint rim,
while the diameter decreases to about 2 arcsec along the tail.  The transition
from the linear segment to the narrower tail is visible as a kink, 2.8\,arcsec
from the head on east side. It is less clearly visible on the west side.

\subsubsection{H$_2$ emission from the surface of knots}

\begin{figure}
\centering
\resizebox{\hsize}{!}{\includegraphics*{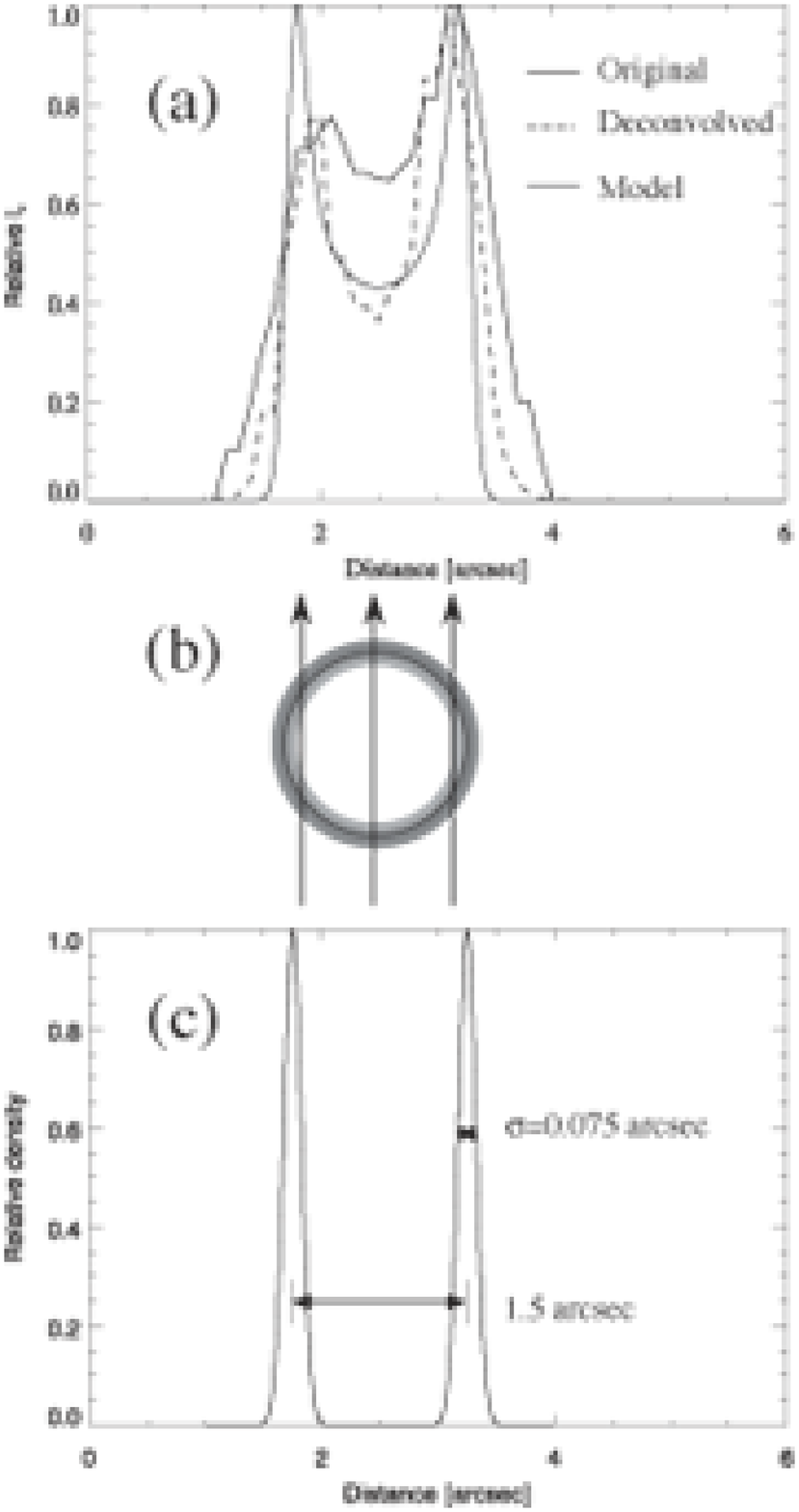}}
\caption{
(a) Cross cut of the observed intensity and at 2.12\,$\mu$m 
and PSF deconvolved profile
along the radial direction in the head. 
The intensities are normalised at the peak.
The location is 1 arcsec away from the brightest tip.
Model profile in panel (a) is to demonstrate hollow shape  in projected image 
if the knot is seen from the central star (panel b) 
with input parameters of the shape (panel c)
\label{Fig:radial}
}
\end{figure}

The deconvolved image (Fig.~\ref{Fig:deconvolved}) shows a limb-brightened
head of the knot.  Within the head there may be an H$_2$ empty region.
Fig.\ref{Fig:radial} shows the radial cross-section of the head 1~arcsec
inside from the brightest point.  The intensity at the mid-point is about 65\%
of the peak in the raw image and about 40\% in the deconvolved image.  To
demonstrate the shape, we modelled the radial profile in two dimensions,
assuming a thin ring with a large hollow area inside (Fig.\ref{Fig:radial}
b and c).  The diameter of the circular ring is 1.5 arcsec.  The
density structure of the ring is assumed to be a Gaussian with a width of
0.075 arcsec. The radial cut in the deconvolved image is reasonably well reproduced
by this model (Fig.~\ref{Fig:radial} top).  The H$_2$ emitting region 
is probably a very thin surface of the knot, except at the tip.

\subsubsection{Comparisons with HST [N{\small{II}}]  and 
[O{\small{III}}]  images} \label{sect:HST}

\begin{figure}
\centering
\resizebox{\hsize}{!}{\includegraphics*{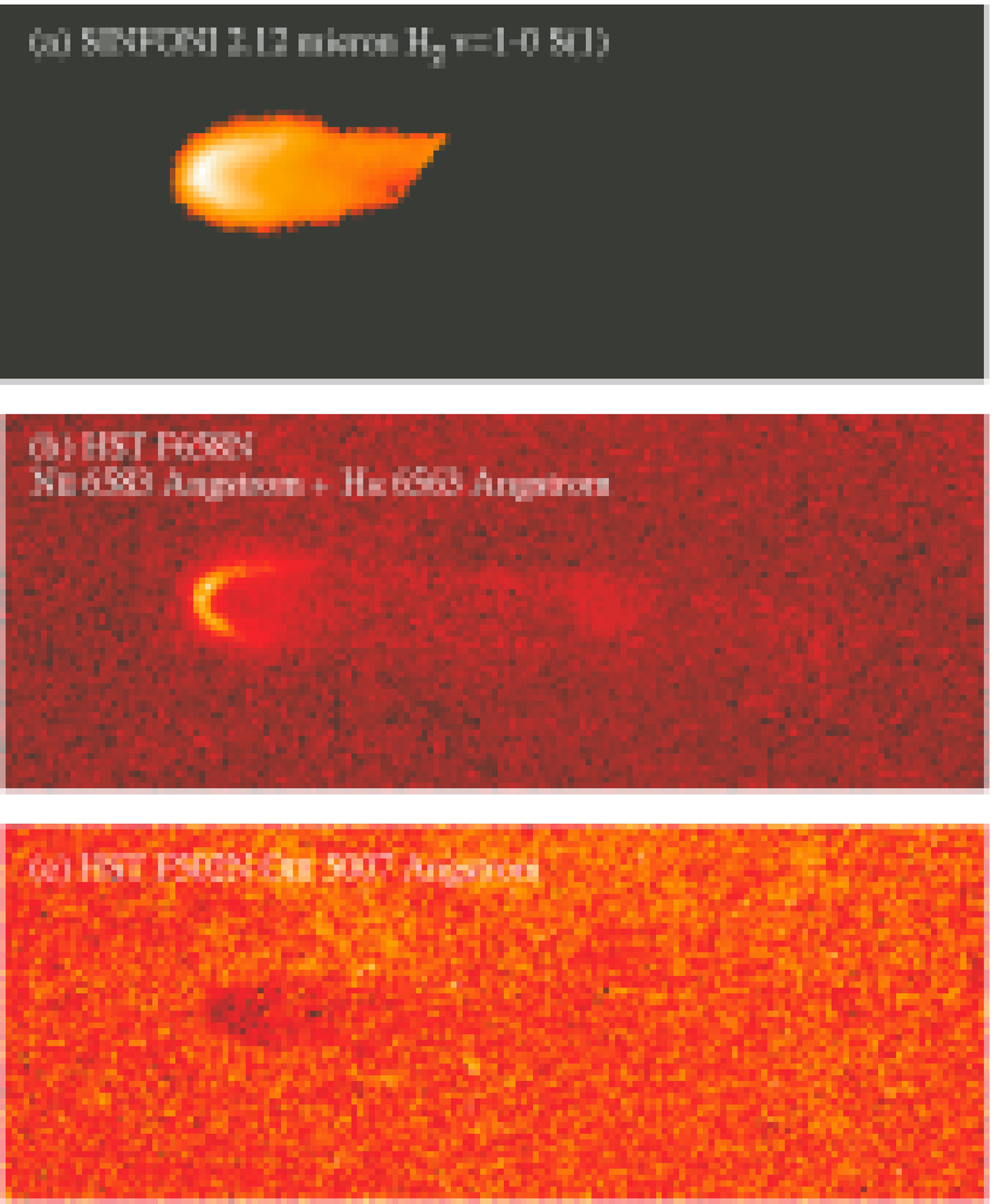}}
\caption{ 
Comparison of the SINFONI 2.12\,$\mu$m image and HST images.
The  SINFONI and HST images were aligned on the tip of the knot. 
Log intensity scale is used for all of the images. The size of the
image is $22.3\times8.5$\,arcsec$^2$
}
\label{Fig:hst}
\end{figure}
\begin{figure}
\centering
\resizebox{\hsize}{!}{\includegraphics*[77,60][371,584]{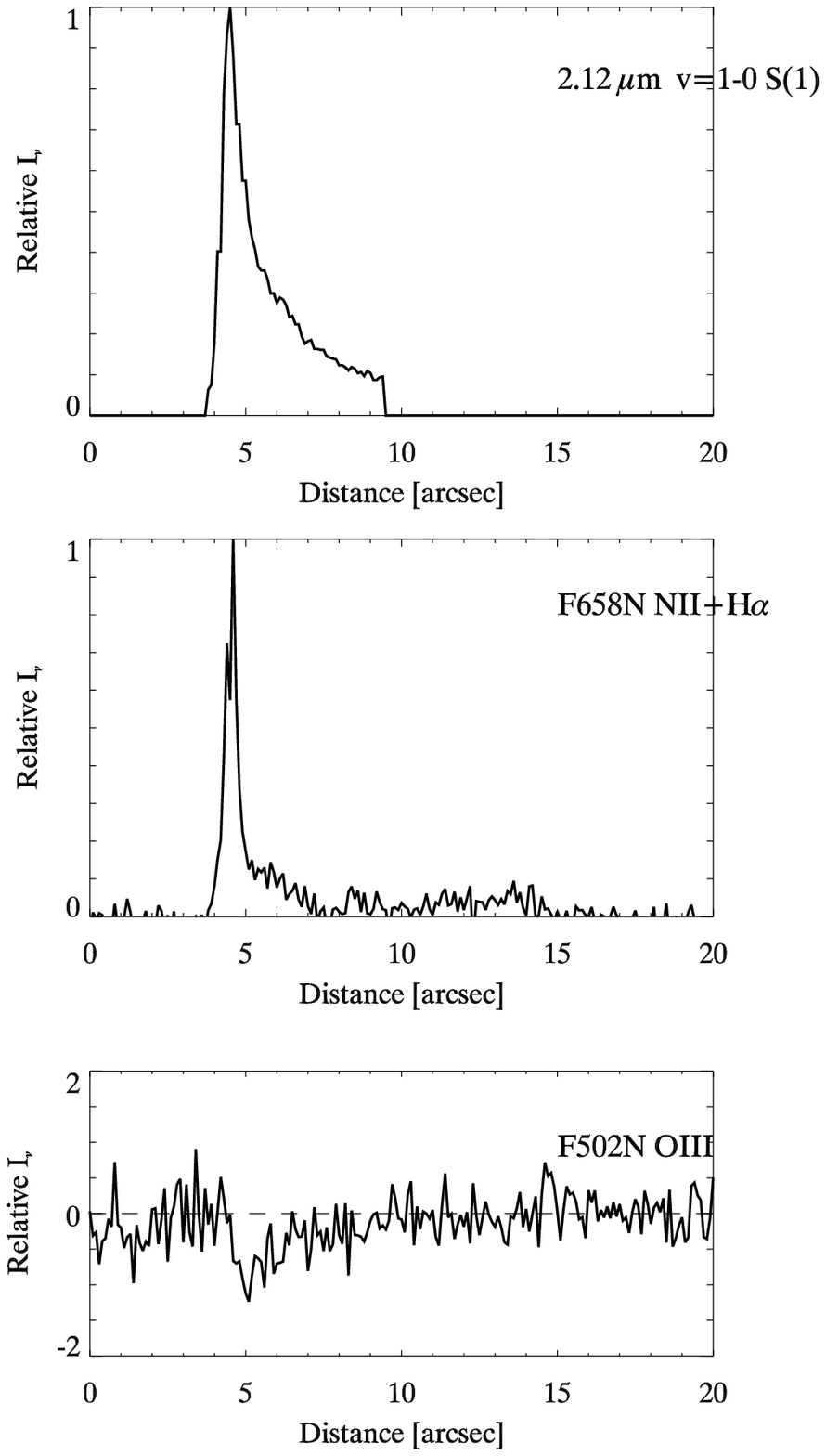}}
\caption{ Cross cut of Fig.~\ref{Fig:hst} along the long axis.
}
\label{Fig:hst-cross}
\end{figure}


Fig.\,\ref{Fig:hst} shows the image of the knot K1 at 2.12~$\mu$m H$_2$ v=1--0
S(1), with a comparison of HST F658N (mainly [N{\small{II}}] and some
contribution of H$\alpha$) and F502N ([O{\small{III}}]) images from
\citet{Odell04}. The precise alignment between SINFONI and HST images is
unknown, and the images were registered such that the tip of the knot is
located at the same place at [N{\small{II}}] and H$_2$ v=1--0 S(1).

The [O{\small{III}}] image shows the knot in absorption; the tip is
slightly off the peak from [N{\small{II}}] image (Fig.~\ref{Fig:hst-cross}).
This has also been  found by \citet{Odell00} for other knots.

A cross section of the H$_2$ v=1--0 S(1) and [N{\small{II}}]+
H{\small{$\alpha$}} images shows that the decay of the intensities towards the
tail is very fast (almost immediate) for the ionized lines, and slower for the
H$_2$ v=1--0 S(1).  The main ionised region is a thin layer at
the tip; the majority of the material remains molecular or neutral.


\subsection{Spectra}

\begin{figure*}
\centering
\resizebox{\hsize}{!}{\rotatebox{90}{\includegraphics*{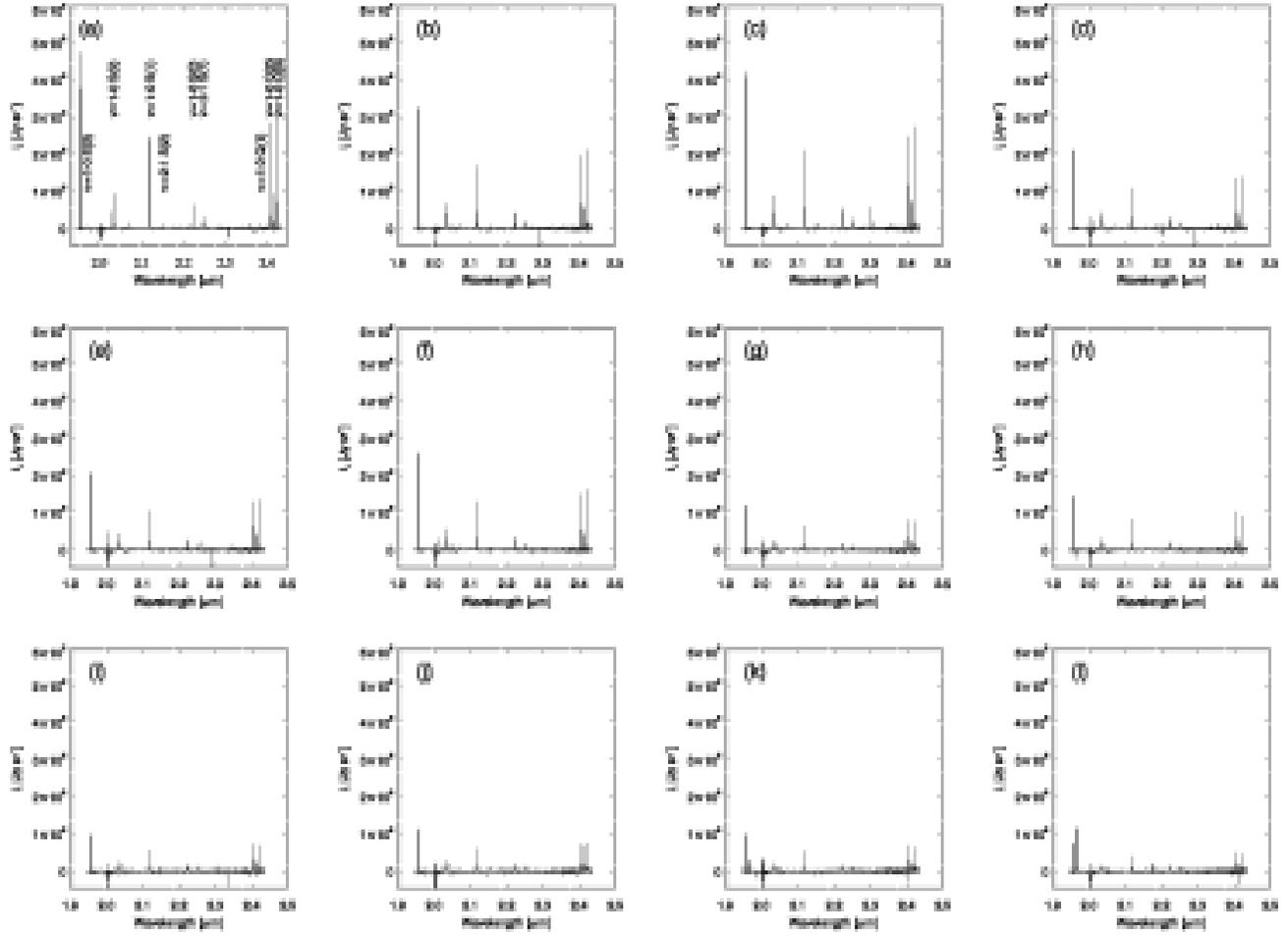}}}
\caption{
Spectra within the knot K1 at twelve 0.675$\times$0.5 arcsec$^2$ areas 
defined in Fig.~\ref{Fig:region_p5g1_250mas}. The y-axis shows the spectral 
intensity from $-0.5\times10^{8}$ to $6\times10^{8}$ Jy\,sr$^{-1}$.
The identifications of the H$_2$ transitions are indicated in panel (a).
\label{Fig:spectra}}
\end{figure*}
\begin{figure}
\centering
\resizebox{\hsize}{!}{\includegraphics*[29,103][278,296]{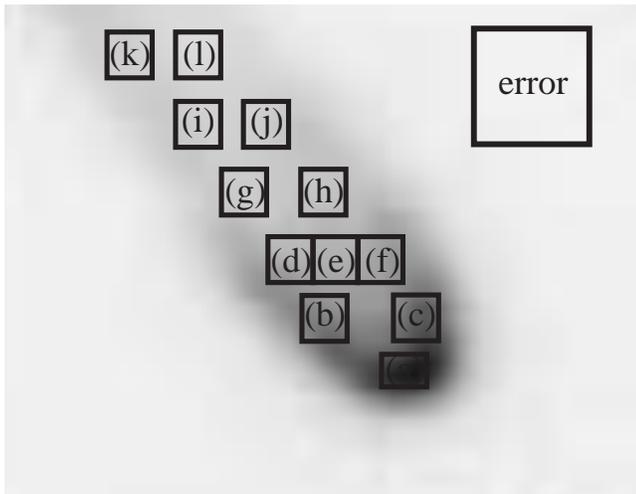}}
\caption{
The twelve regions shown by boxes on the 2.12 $\mu$m image. 
The spectra are averaged within that area, which corresponds to the diagram 
in Fig.~\ref{Fig:h2ratio1}. The
right top box is the area used to estimate the error of the intensities.
\label{Fig:region_p5g1_250mas}}
\end{figure}
\begin{figure*}
\centering
\resizebox{\hsize}{!}{\rotatebox{90}{\includegraphics*{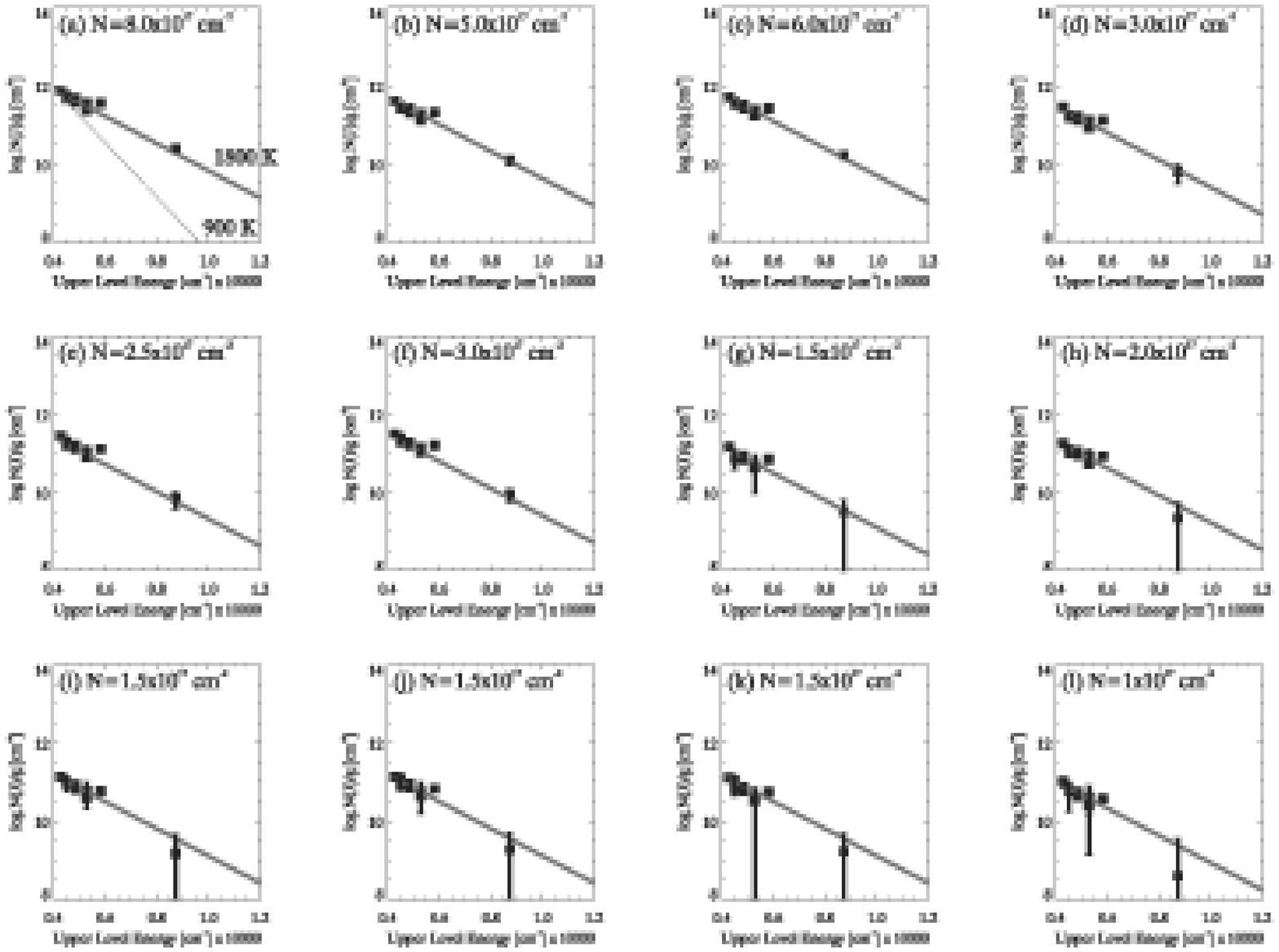}}}
\caption{ The energy diagrams for area (a)--(l) defined in
Fig. ~\ref{Fig:region_p5g1_250mas}; the column density for a given J upper
level divided by the statistical weight as a function of upper state energy in
cm$^{-1}$.  The square shows the measured energy levels.  The thick lines show
the LTE case for 1800~K and for column densities as given in the diagrams.
For a comparison, we plot the 900~K LTE line, as derived for knots in the outer
region of the nebula \citep{Cox98} in (a).  The column density for the 900\,K
line is $8\times10^{18}$cm$^{-2}$.  This temperature cannot fit the H$_2$ line
intensities measured in the knot K1.
\label{Fig:h2ratio1}}
\end{figure*}

Fig.~\ref{Fig:spectra} shows the spectra of the knot K1 at twelve regions
indicated in Fig.~\ref{Fig:region_p5g1_250mas}.  Within each region of 5 by 5
pixels (except region (a) which has 5 by 4 pixels), 
the SINFONI data have been averaged.  At most twelve H$_2$ lines
(three of them with low signal-to-noise ratio) are detected by SINFONI. The
intensities are summarised later in Table~\ref{table-ratio}.  The strongest
detected line is the 1.95~$\mu$m H$_2$ v=1--0 S(3), and the 2.12~$\mu$m H$_2$ v=1--0
S(1) and 2.4~$\mu$m v=1--0 Q-branches are also strong. The H$_2$ lines are
unresolved at the resolution of R=5090. The 2.07\,$\mu$m v=2--1 S(3) and
2.15\,$\mu$m v=2--1 S(2) 2.20\,$\mu$m v=3-2 S(3) lines are only marginally
detected.

Several high transition lines within this wavelength range
are not detected, such as the 2.00 $\mu$m v=2--1 S(4), 2.38 $\mu$m v=3--2 S(1).  
The Br$\gamma$ line is not detected.

\begin{table*}
  \caption{
  Intensities of H$_2$ lines at area (a)--(c) and ratios with respect to
  the v=2--1 S(1) line. 
   \label{table-ratio}}
\begin{center}
 \begin{tabular}{llrrr llrrrrrccccccccc}
  \hline
Wav & Transition &  \multicolumn{2}{c}{Upper state energy} & Statistical weight & \multicolumn{3}{c}{Intensity $I_{\nu}$$\times10^7$} 
& Err$^{\S}$ & \multicolumn{3}{c}{Obs ratio$^{\P}$}  \\
& & & & & (a) & (b) & (c) & & (a) & (b) & (c) & \\
$\mu$m & & K & cm$^{-1}$ & & \multicolumn{3}{c}{W\,m$^{-2}$sr$^{-1}$} &  &  \\ \hline
   1.958 & v=1--0 S(3)$^{\dagger}$ &  8365 &  5813    & 33   & ~4.10 &  2.35 &  2.96 & 0.10 & 220  &   226  &  223 \\
   2.004 & v=2--1 S(4)                        & 14764 & 10262  & 13   & $<$0.40$^{\S}$ \\
   2.034 & v=1--0 S(2)                        &   7584 &    5271  &   9   & ~0.67 &  0.37 &  0.49 & 0.04 & 36   &      35  &   36  \\
   2.073 & v=2--1 S(3)                        & 13890 &   9654   & 33  & ~0.08$^{\ddag}$ &  &  & 0.09 & 4 &  &  \\
   2.128 & v=1--0 S(1)                        &   6956 &   4834   & 21  & ~1.86 &  1.04 &  1.32 & 0.07 & 100 &   100  & 100  \\
   2.154 & v=2--1 S(2)                        & 13150 &   9139   &  9   & ~0.07$^{\ddag}$ &  & & 0.08  &     3  &          &        \\
   2.201 & v=3--2 S(3)                        & 19086 & 13265   & 33 & ~0.02$^{\ddag}$ & & & 0.03\\
   2.224 & v=1--0 S(0)                        &   6471 &    4497   &   5 & ~0.42 &  0.24 &  0.31 & 0.06 &   22  &    23   &  23    \\
   2.248 & v=2--1 S(1)                        & 12550 &    8722  & 21 & ~0.18 &  0.09 &  0.12 & 0.04  &   9  &       9  &     9    \\
   2.386 & v=3--2 S(1)                        & 17818 &   12384 & 21 & $<$0.09$^{\S}$ \\
   2.407 & v=1--0 Q(1)$^{\dagger}$ & 6149 &  4273      &   9 & ~1.84 &  1.06 &  1.30 & 0.23 & 99 &   102  &   97   \\
  2.413 & v=1--0 Q(2)$^{\dagger}$  & 6471 &  4497      &   5 & ~0.55 &  0.29 &  0.39 & 0.12 &   29 &      28  &   29   \\
  2.424 & v=1--0 Q(3)$^{\dagger}$  & 6956 &  4834      & 21 & ~1.85 &  1.03 &  1.27 & 0.14 &  99 &   99  &   96   \\
  2.438 & v=1--0 Q(4)$^{\dagger}$  & 7586 &  5272      &   9 & ~0.57 &  0.31 &  0.38 & 0.18 &   30 &      30  &   28  \\
\hline
\end{tabular}
\end{center}
$^{\P}$ Line ratio with respect to 2.128\,$\mu$m v=1--0 S(1) line \\
$^{\S}$3$\sigma$ of the noise level of the intensity  in the error region indicated in Fig.\,\ref{Fig:region_p5g1_250mas}. \\
$^{\dagger}$These lines have a large ($\sim$30\%) error in relative intensity 
calibration due to the influence of H$_2$O in the terrestrial atmosphere.
They are excluded from the comparison with models. \\
$^{\ddag}$ Marginal detections \\
\end{table*}

\subsubsection{Energy diagram}

Fig.\,~\ref{Fig:h2ratio1} shows the energy diagram of the H$_2$ lines for
several regions in the knot indicated in Fig.\,~\ref{Fig:region_p5g1_250mas}.
Einstein coefficients from \citet{Turner77} are used.  The slopes in the
diagrams show that the line intensities follow a $T= 1800~$K Local
Thermodynamic Equilibrium (LTE) distribution.  At the bright area (a--c), the
observed line intensities follow the 1800~K LTE line up to the upper energy of
8000~cm$^{-1}$.  The systematic uncertainty in the absolute H$_2$ intensity
(up to 50 per cent) affects the column density, but not the excitation
temperature which is determined by the slope.

\subsection{Line ratio map} \label{sect:lineratiomap}

\begin{figure*}
\centering
\resizebox{\hsize}{!}{\includegraphics*[0,110][1138,686]{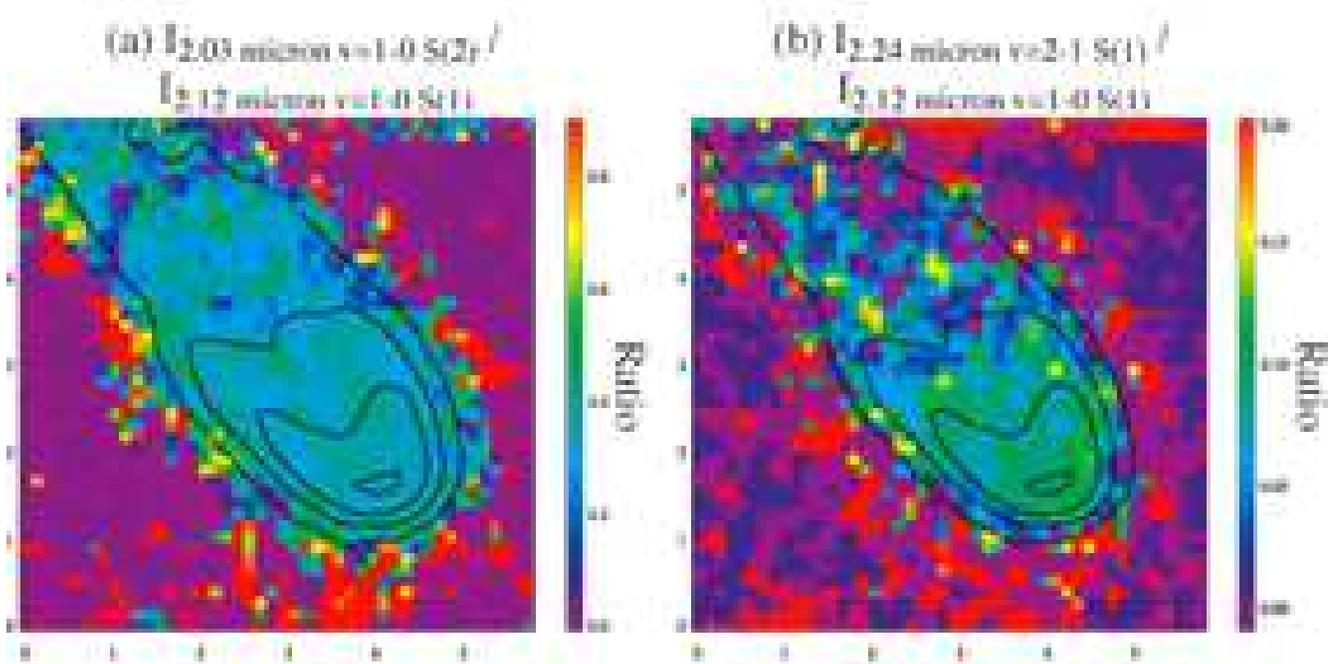}}
\caption{
The line ratio maps of the cometary knot K1. The contour shows the surface 
brightness of 2.12~$\mu$m H$_2$ v=1--0 S(1) line at $0.2\times10^{-7}$, 
$0.5\times10^{-7}$, $1\times10^{-7}$, $2\times10^{-7}$ W\,m$^{-2}$\,sr$^{-1}$.
\label{Fig:ratio_p5g1_250mas}
}
\end{figure*}

Fig.\,~\ref{Fig:ratio_p5g1_250mas} shows the line ratio maps within the knot.
Fig.\,~\ref{Fig:ratio_p5g1_250mas} (a) maps the rotational temperature
variations within the knot, represented by the 2.03~$\mu$m v=1--0 S(2) and
2.12~$\mu$m v=1--0 S(1) lines.  This rotational temperature is uniform within
the errors throughout the knot.  This uniform rotational temperature is
expected from the energy diagram (Fig.\,~\ref{Fig:h2ratio1}), because all of
the line intensities follow 1800~K LTE in any area.

The vibrational temperature seems to vary within the knot but this is within the
uncertainty.  Fig.\,~\ref{Fig:ratio_p5g1_250mas} (b) shows the line ratio of
2.24~$\mu$m H$_2$ v=2--1 S(1) and 2.12~$\mu$m H$_2$ v=1--0 S(1).  This is the
only combination to obtain the vibrational temperature within our observed
data.  The line ratio tends to be highest at the tip ($\sim 0.1$) and to
decrease towards the tail, although a careful treatment of v=1--0 S(0) line is
needed.

\section{Discussion}

\subsection{Excitation diagrams and temperatures}

The H$_2$ excitation diagrams are fitted by a single LTE temperature of
1800~K. This is much higher than \citet{Cox98} who obtained 900\,K by fitting
pure-rotational lines up to the S(7) transition.  The region observed by
\citet{Cox98} is located at the western rim of the nebula, and is
5--6\,arcmin away from the central star.  The pixel size of their ISOCAM data
was 6\,arcsec and emission from multiple knots contributed to each single
pixel.  In contrast, our target is a single, isolated knot located at the
innermost region of the flock of knots (2.5\,arcmin from the central star).

\citet{Odell07} also obtained an excitation temperature of 988\,K
from 5--15\,$\mu$m spectra in the outer part of the nebula.  
\citet{Odell07}
state that the distance of that slit positions is similar to that of \citet{Cox98}.
Although they do not provide detailed information on the slit positions
and observing mode for their two Spitzer spectra, the Spitzer archive can provide
this, suggesting their two slit positions are 5.6 and 4.2\,arcmin from the central star,
and the slit size is 3.6$\times$57\,arcsec$^2$.
Further three spectra at different slit positions are available in archive, but these
do not fit the description of the data in \citet{Odell07}.

 The difference in measured temperatures indicates that the excitation
temperature of H$_2$ is not uniform within the Helix nebula: the H$_2$
molecules reach higher temperature within the inner region.  From about
2.5\,arcmin to 5\,arcmin away from the central star, the excitation
temperature decreases from 1800\,K to 900--1000\,K.
This provides evidence for temperature variations within the nebula.

\citet{Hora06} show that the [3.6]$-$[4.5] vs [4.5]$-$[8.0] colour varies
within the nebula. All three bands are dominated by H$_2$ lines.  Knots
located at the inner rim of the main nebula are brighter in the 4.5~$\mu$m
band. The strongest expected line in this band is the 4.69\,$\mu$m H$_2$ 0--0
S(9), whose upper energy is the highest among the dominant H$_2$ lines within
the IRAC filters.  The colour-colour diagram of \citet{Hora06} also suggests a
higher excitation temperature in the inner region, and global variation in the
excitation temperature in the entire nebula.
Analysis of spectra with several slit positions
within a nebula shows that the H$_2$ temperature is not uniform within a PN
\citep{Hora99, Davis03}.  Excitation temperature variations appear
normal within a single PN.

Furthermore, H2 excitation temperatures in PNe have been found as high as 2000 K (Hora et al. 1999; Davis et al. 2003) and lower than 1000K 
\citep{Cox98, Bernard-Salas05, Matsuura05}.
Variations in excitation temperature both between PNe and within a single PN appear to be normal

\subsection{Column density}

The H$_2$ line intensities are fitted using column densities in the range
$1\times10^{17}$--$8\times10^{17}$\,cm$^{-2}$.  This range is smaller than the
\citet{Cox98} measurements of $3\times10^{18}$\,cm$^{-2}$.  Our target is an
isolated knot, while the low spatial resolution of the ISOCAM data almost
 included multiple knots within the field of view, hence the H$_2$
intensity is much higher \citep{Speck02}. 

The column density of H$_2$ can be converted to a hydrogen mass of $2\times
10^{-8} M_{\odot}$.  Here we assume the knot has a column density of $4\times
10^{17}$~cm$^{-2}$ (i.e. to emit average intensity over 2$\times$2~arcsec$^2$
at the excitation temperature of 1800\,K), the distance is 219~pc, the
dimension of the knot is 2$\times$2~arcsec$^2$ and the density is uniform
within this area.  The estimated hydrogen mass is a factor of 1000 less than
the estimate of \citet{Odell96}, who used the extinction of the [O{\small{II}}] line
as a mass tracer.  Our infrared H$_2$ lines trace only highly excited H$_2$, as
expected from the upper state energy ($>$6000\,K). Colder H$_2$ gas is
inefficient at emitting these lines \citep{Cox98}.  

In order to measure the mass of cold H$_2$ gas more directly,
 we need observations at UV wavelengths, where H$_2$ lines could be found in absorption. 
However, this approach may be compromised by the contribution to H$_2$ absorption by the ISM
Adopting the hydrogen mass of \citet{Odell96} with a comparison of our estimated H$_2$ mass
indicates that H$_2$ is heated only at
the surface, but is in fact present throughout the knot. This favours a
scenario where the detected H$_2$ was already present within the knot,
i.e., it is not necessary to assume that it formed from chemical reactions from
atomic/neutral hydrogen within this surface. However, this argument is based
on an expected correlation between dust extinction and H$_2$. A direct
detection is required to confirm the presence of a large, cold H$_2$
reservoir.

\subsection{Excitation mechanisms of molecular hydrogen}

The cometary knot K1 emits a 1--0 S(1) line intensity of
$2\times10^{-7}$\,W\,m$^{-2}$\,sterad$^{-1}$. The adopted uncertainty is a
factor of two. The 2--1/1--0 S(1) line ratio is $\sim0.1\pm0.02$ at the tip,
using the uncertainty on the response correction.

 \citet{Burton92} calculated the line intensities of 1--0 S(1) and 2--1 S(1) lines
for J-type shocks, C-type shocks and in PDR.  We found that within their model
the measured line
intensities and ratios can be fit equally well under several model conditions.

\begin{itemize}
\item
A photo-dissociation region (PDR) heated by
UV radiation where the density is 
$n=10^{6}$\,cm$^{-3}$ and the UV strength is $G_0=1.2\times10^4$
(within the range of $G_0=$1--1.2$\times10^4$)
\item 
C-type shocks with upstream density $n_0=10^{4}$\,cm$^{-3}$ and shock velocity 
$v_s\sim27$\,km\,s$^{-1}$ (within the range of 27--28\,km\,s$^{-1}$).
\item
J-type shocks  $n_0=10^{6}$\,cm$^{-3}$ and shock velocity 
$v_s\sim9$\,km\,s$^{-1}$(9--10\,km\,s$^{-1}$).
\end{itemize}

Although there is some uncertainty in absolute intensity,
the line ratio is the strongest constraint on the above conditions.

%

\citet{Meaburn98} estimated the molecular hydrogen density of another knot to
be $8.9\times10^{5}$\,cm$^{-3}$ using the extinction in the [O{\small{III}}]
5007\,\AA\, line.  This knot appears brighter than K1, and larger in radius
when observed in the [N{\small{II}}] line.  We follow their method to estimate
the density of the knot K1.  The absorption at [O{\small{III}}] is 0.73 with
respect to the continuum (Sect.\ref {sect:HST}).  This corresponds to an
extinction coefficient $c=0.15$, $E(B-V)=$0.14~mag, and an H$_2$ column
density of $4.0\times10^{20}$\,cm$^{-2}$.  If the diameter of the knot is 1.5
\,arcsec, an H$_2$ number density of $8\times10^4$\,cm$^{-3}$ is obtained for a
distance of 219\,pc.  The derived H$_2$ density is consistent for PDR
model of H$_2$ excitation in \citet{Burton92}.

The FUV (6--13.6 eV) flux at the knot K1 is about $G_0=8 $, where $G_0$ is 
the FUV radiation measured in units of the \citet{Habing68} flux. 
This is based on \citet{Su07}'s estimate assuming a luminosity
of the central star of  76\,$L_{\sun}$ and a distance of 219\,pc.
The required UV radiation in \citet{Burton92}'s model is 
$G_0>1\times10^{4}$, which is an order of magnitude higher 
than the estimated UV radiation field strength at the knot K1.
We used the UCL-PDR model \citep{Bell05} to calculate the conditions
independently.  At an UV radiation field of $G_0=8$, we obtain a temperature
below 100\,K.  
\citet{Roellig07} compare benchmark calculations of independent PDR codes,
including the UCL-PDR model and \citet{Tielens85}'s model that was
incorporated in \citet{Burton92}'s H$_2$ model.  They find consistent gas
temperatures among PDR models in their benchmark calculations.  These results
suggest that the PDR models can reproduce the measured LTE gas temperature
(1800 K) of the H$_2$ lines in the knot K1 only, if the UV
radiation strength is increased.  A similar conclusion is derived by \citet{Cox97} and
\citet{Odell05}.  The value for $G_0$ is defined as the flux between
6--13.6\,\AA\, a narrow-band UV radiation field, adopted for interstellar
radiation field. If we consider the continuous stellar spectrum of the central
star, the UV flux increases by a factor of 10 \citep{Odell07}. However,
this remains a factor of 250 short of the required flux.

The speed of ambient gas
for C-type shocks is $v_s\sim27$\,km\,s$^{-1}$ to fit
the H$_2$ line intensities and line ratios, from theoretical work by
\citet{Burton92}.  The required velocity can vary by $\sim\pm10$\,km\,s$^{-1}$
depending on the assumed magnetic field strength and the iron fraction.  The
[He{\small{II}}] and [O{\small III}] lines show an expansion velocity of
$13$\,km\,s$^{-1}$ near the central star \citep{Meaburn05}. Within an
expanding nebula, hydrodynamic effects will cause significant velocity
gradients \citep{Schoenberner05}, as are observed in PNe \citep{Gesicki03}. 
The overpressure of the ionised region dominates, and (if present) the pressure from
the inner, hot bubble might be added. The same processes will occur for each knot at the
ionized, facing edge. \citet{Meaburn98, Meaburn05} postulate turbulent velocities 
$\ge10$\,km\,s$^{-1}$ (i.e. larger than the sound speed). 
A C-type shock velocity could be associated with such motion.  The
gas density of PNe is typically $n_0 \sim 10^4$\,cm$^{-3}$.  This is
consistent with the required upstream density.  
C-type shock excitation of H$_2$ is possible. Stronger observational constraints on the velocity
structure of the knot would be helpful.

The presence of magnetic fields has been reported in several young or
pre-planetary nebulae using sub-mm polarization \citep{Sabin07}.  They find a
magnetic field of $\sim$1mG at $5 \times 10^{16}\,$cm from the central star in
other PNe.  The Helix knots are located 10 times further from the star than
their measured location.  A dipole field would decay as $r^{-3}$: this would
leave a negligible field in the Helix. A solar-type field decays as $r^{-2}$,
and a frozen toroidal field \citep[as favoured by][]{Sabin07} may decay slower
with radius.  The formation of a knot may strengthen after its embedded magnetic
field.  The field required for the C-type shock is plausible within the knot,
compared to the stronger fields detected in more compact shells. The
inter-knot medium is likely to show a weaker field.

An upstream (wind) density of $10^{6}$\,cm$^{-3}$ is required for the J-type shock
model.  Although the density within knots themselves are recorded at
$10^6$\,cm$^{-3}$ \citep{Odell96, Meaburn98}, it is unlikely that the upstream
region has such a high density.

\section{Conclusions}

We have investigated the detailed structure of
a single knot close to the inner edge of the main ring of the Helix nebula.
 We find that the rotational-vibrational temperature of H$_2$ is as high as 1800\,K for
this innermost cometary knot. The rotational temperature is uniform within the
knot, 
and the vibrational temperature appears to follow the same distribution 
except for a possible decrease towards the tail.
The derived
temperature is much higher than previously measured (900\,K) for knots in the
outer region.  The excitation temperature changes with radial distance from
the central star.  The studied knot has a wide head, with the H$_2$
distributed in a crescent.

We examine the possible molecular hydrogen excitation mechanisms.
Based on the line intensities and ratios, C-type shocks can provide a
fit using plausible local conditions, althought the required velocity is slightly higher
than the observed one. 
This excitation requires the presence of an as-yet
undetected magnetic field. A J-type shock model does not fit.
We also tested the possibility that H$_2$ lines are excited in PDR, 
but PDR models have difficulties to reproduce the
very high temperature, unless the UV field is increased by two orders
of magnitude over the stellar radiation field. 
A further test of the models will require observations of the velocity
structure of the head and tail. We also lack a synthesis model combining
collisional and UV excitation, with hydrodynamical interactions.

\section{acknowledgements}
We appreciate technical support from ESO staff during the observations and the
data-analysis.  M.M. appreciates encouragement from Prof. Arimoto for this
study. MM is grateful for hospitality at the UCL and SAAO during the visits.
A discussion with Dr. M. Cohen in early stage of this research was very useful.

\label{lastpage}
\end{document}